\documentclass[10pt]{article}
\usepackage{epstopdf}

\usepackage{amsmath}
\usepackage{amssymb}

\usepackage{graphicx}

\usepackage{cite}

\usepackage{color} 

\usepackage{epsfig}
\usepackage{algorithm}
\usepackage[noend]{algorithmic}
\usepackage{algorithmic}
\usepackage{subfigure}
\usepackage{amssymb}
\usepackage{amsmath}
\usepackage{stmaryrd}
\usepackage{url}

\usepackage[labelfont=bf,labelsep=period,justification=raggedright]{caption}

\makeatletter
\renewcommand{\@biblabel}[1]{\quad#1.}
\makeatother

\date{}

\begin{document}
\epstopdfsetup{outdir=./}
\begin{flushleft}
{\Large
\textbf{Science vs Conspiracy: collective narratives in the age of (mis)information}
}
\\
Alessandro Bessi$^{1,2}$
Mauro Coletto$^{2}$,
George Alexandru Davidescu$^{2}$,
Antonio Scala$^{2,3}$,
Guido Caldarelli$^{2,3}$,
Walter Quattrociocchi$^{2,4}\ast$
\\
\bf{1} IUSS Institute for Advanced Study, Piazza della Vittoria 5, 27100 Pavia, Italy
\\
\bf{2} Laboratory of Computational Social Science, Networks Dept IMT Institute for Advanced Studies Lucca, 55100 Lucca, Italy
\\
\bf{3} ISC-CNR Uos "Sapienza", 00185 Roma, Italy
\\
\bf{4} Laboratory for the Modeling of Biological and Socio-technical Systems, Northeastern University, Boston, MA 02115 USA
\\$\ast$ Corresponding Author E-mail: walter.quattrociocchi@imtlucca.it
\end{flushleft}

\section*{Abstract}
The large availability of user provided contents on online social media facilitates people aggregation around common interests, worldviews and narratives. However, in spite of the enthusiastic rhetoric about the so called {\em wisdom of crowds}, unsubstantiated rumors -- as alternative explanation to main stream versions of complex phenomena -- find on the Web a natural medium for their dissemination.
In this work we study, on a sample of 1.2 million of individuals, how information related to very distinct narratives -- i.e. main stream scientific and alternative news -- are consumed on Facebook.  
Through a thorough quantitative analysis, we show that distinct communities with similar information consumption patterns emerge around distinctive narratives.
Moreover, consumers of alternative news (mainly conspiracy theories) result to be more focused on their contents, while scientific news consumers are more prone to comment on alternative news.
We conclude our analysis testing the response of this social system to 4709 troll information -- i.e. parodistic imitation of alternative and conspiracy theories. 
We find that, despite the false and satirical vein of news, usual consumers of conspiracy news are the most prone to interact with them.


\section*{Introduction}
The growth of knowledge fostered by an interconnected world together with the unprecedented acceleration of scientific progress has exposed the society to an increasing level of complexity to explain reality and its phenomena. At the same time, a shift of paradigm in the production and consumption of contents has occurred, utterly increasing the amount and heterogeneity of information available to users. 
Everyone on the Web can produce, access and diffuse contents by actively participating in the creation, diffusion and reinforcement of different and heterogeneous narratives.
The large heterogeneity of information fostered the aggregation of people around common interests, worldviews and narratives.

Recently in \cite{Mocanu2014} it has been shown that online unsubstantiated rumors -- such as the link between vaccines and autism, the global warming induced by chem-trails or the secret alien government -- and main stream information -- such as scientific news and updates -- reverberate in a comparable way. This pervasiveness of unreliable contents might lead to mix up unsubstantiated stories with their satirical counterparts -- e.g. the presence of sildenafil-citratum (the active ingredient of Viagra\texttrademark) \cite{sildenafil} in chem-trails or the anti hypnotic effects of lemons (more than 45000 shares) \cite{limoni1,limoni2}.
In fact, there are very distinct groups, namely {\em trolls}, building Facebook pages as a caricatural version of conspiracy news. Their activities range from controversial comments and posting satirical contents mimicking alternative news sources, to the fabrication of purely fictitious statements, heavily unrealistic and sarcastic. Not rarely, these memes became viral and were used as evidence in online debates from political activists \cite{forconi_cirenga2013}.

The vast majority of unsubstantiated rumors on the web is related to conspiracy stories and find on the Web a natural medium for their dissemination.
Narratives grounded on conspiracy theories tend to reduce the complexity of reality and are able to contain the uncertainty they generate \cite{byford2011conspiracy,finerumor,hogg2011extremism}.
They are able to create a climate of disengagement from mainstream society and from officially recommended practices \cite{bauer1997resistance} -- e.g. vaccinations, diet, etc.

Despite the enthusiastic rhetoric about the {\em wisdom of crowds} \cite{surowiecki2005wisdom,WelinderBBP10} -- i.e. collective decisions are wiser than judgements of single individuals -- the role of socio-technical system in enforcing informed debates and their effects on the public opinion still remain unclear.
The World Economic Forum listed massive digital misinformation as one of the main risks for modern society \cite{Davos13} because, even if the lifetime of false information is short and it is generally unlikely to result in severe real-world consequences, it is conceivable that a false rumor spreading virally through social networks might impact the public opinion before being effectively corrected. 
A multitude of mechanisms animates the flow and acceptance of false rumors, which in turn create false beliefs that are rarely corrected once adopted by an individual \cite{Garrett2013,Meade2002,koriat2000,Ayers98}.
The process of acceptance of a claim (whether documented or not) may be altered by normative social influence or by the coherence with the system of beliefs if the individual  \cite{Zhu2010,Loftus2011}.  
A large body of literature addresses the study of social dynamics on socio-technical systems from social contagion up to social reinforcement \cite{khosla2014makes,Onnela2010,Ugander2012,Lewis2012,Mocanu2012,Adamic05thepolitical,kleinberg2013analysis,eRep,QuattrociocchiCL11,QuattrociocchiPC09,Bond2012,BorgeHolthoefer:2011p5170,Centola2010,Castellano2007,Quattrociocchi2014,Bennaim2003,adamic2013,hannak-2012-weather,Cheng:2014}.

In this work we target consumption patterns of information supporting distinct narratives. In particular, focusing on the Italian context and helped by pages very active in debunking unsubstantiated rumors (see acknowledgment section), we build an atlas of scientific and alternative information sources (see supporting material for further details) on Facebook.
On the one hand, conspiracists tend to explain significant social or political aspects with plots conceived by powerful individuals or organizations controlling main stream media. Their arguments can sometimes involve the rejection of science and invoke alternative explanations to replace scientific evidence. For instance, people who reject the link between HIV and AIDS generally believe that AIDS was created by the U.S. Government to control the African American population \cite{Bogart05,Kalichman09}. On the other hand, scientific diffusion is trying to inform users of scientific progress as well as about scientific methodology and rational thinking to deal with the concept of complexity.

Such a context represents an unprecedented possibility to study and characterize with a high level of resolution the relation between content selection, information consumption, beliefs formation and revision, and their effect in shaping communities of interest where information are supporting distinctive and someway opposite narratives.

Our dataset contains 271,296 post created by 73 Facebook pages. 
Pages are classified according to the kind of information disseminated and their self description in conspiracy news -- alternative explanations of reality aiming at diffusing contents neglected by main stream information -- and scientific news.

Notice that it is not our intention claiming that conspiracist and alternative information are necessarily false. Our focus is on how communities formed around different information and narratives interact and consume information.
To do this we account for user interaction with respect to pages public posts -- i.e. likes, shares, and comments. Each of these actions has a particular meaning \cite{Ellison2007,Joinson:2008,Viswanath:2009}. A {\em like} stands for a positive feedback to the post; a {\em share} expresses the will to increase the visibility of a given information; and {\em comment} is the way in which online collective debates take form around the topic promoted by posts. 
Comments may contain negative or positive feedbacks with respect to the post.

Our analysis starts with an outline of information consumption patterns and the community structure of pages according to their common users. 
We label polarized users -- users which their like activity (positive feedback) is almost (95\%) exclusively on the pages of one category -- and find similar interaction patterns on the two communities with respect to their contents.
Moreover, we measure their commenting activity on the opposite category finding that polarized users of alternative news are more focused on posts of their community and that they are more oriented to the diffusion -- i.e. they are more likely to like and share posts from conspiracy pages. On the other hand, usual consumers of scientific news result to be less committed in the diffusion and more prone to comment on conspiracy pages. 
Finally, we test the response of polarized users to the exposure to 4709 satirical and demential version of conspiracy stories finding that, out of 3888 users labeled on likes and 3959 on comments, the most of them are usual consumers of alternative stories (80.86\% of likes and 77.92\% of comments).

\section*{Methods}
\subsection*{Ethics Statement}
The entire data collection process has been carried out exclusively through the Facebook Graph API \cite{fb_graph_api}, which is publicly available, and for the analysis (according to the specification settings of the API) we used only public available data (users with privacy restrictions are not included in the dataset). The pages from which we download data are public Facebook entities (can be accessed by anyone). User content contributing to such pages is also public unless the user's privacy settings specify otherwise and in that case it is not available to us.

\subsection*{Data collection}

The debate around relevant social issues spreads and persists over the web, leading to the emergence of unprecedented social phenomena such as the massive recruitment of people around common interests, ideas or political visions.
We defined the space of our investigation with the help of Facebook groups very active in the debunking of conspiracy theses.
The resulting dataset is composed of 73 public pages divided in scientific and conspiracist news for which we download all the posts (and their respective users interactions) in a timespan of 4 years (2010 to 2014).
The exact breakdown of the data is presented in Table \ref{tab:data_dim}. 
The first category includes all pages diffusing alternative information sources -- pages which disseminate controversial information, most often lacking supporting evidence and sometimes contradictory of the official news (i.e. conspiracy theories). 
The second category is that of scientific dissemination including scientific institutions and scientific press having the main mission to diffuse scientific knowledge.

\begin{center}
\begin{table}[!h]
\centering
\begin{tabular}{l|c|c|c|c}
\bf {  }  & \bf {Total} & \bf {Scientific News} & \bf {Alternative News}  \\ \hline 
Pages & $ 73 $ & $ 34 $ & $ 39 $ \\
Posts & $ 271,296 $ & $ 62,705 $ & $ 208,591 $ \\
Likes & $ 9,164,781 $ & $ 2,505,399 $ & $ 6,659,382$  \\
Comments & $1,017,509 $ & $ 180,918 $ & $ 836,591 $\\
Unique Likes & $ 1,196,404 $ & $ 332,357 $ & $ 864,047 $\\
Unique Comments & $ 279,972 $ & $ 53,438 $ & $ 226,534 $\\
\end{tabular}
  \caption{ \textbf{Breakdown of Facebook dataset.} The number of pages, posts, comments and likes for alternative and scientific news.}
  \label{tab:data_dim}
\end{table}
\end{center}

\section*{Results and discussion}

\subsection*{Consumption patterns on scientific and alternative news}
\begin{figure}[H]
 \centering
       \includegraphics[width=.60\textwidth]{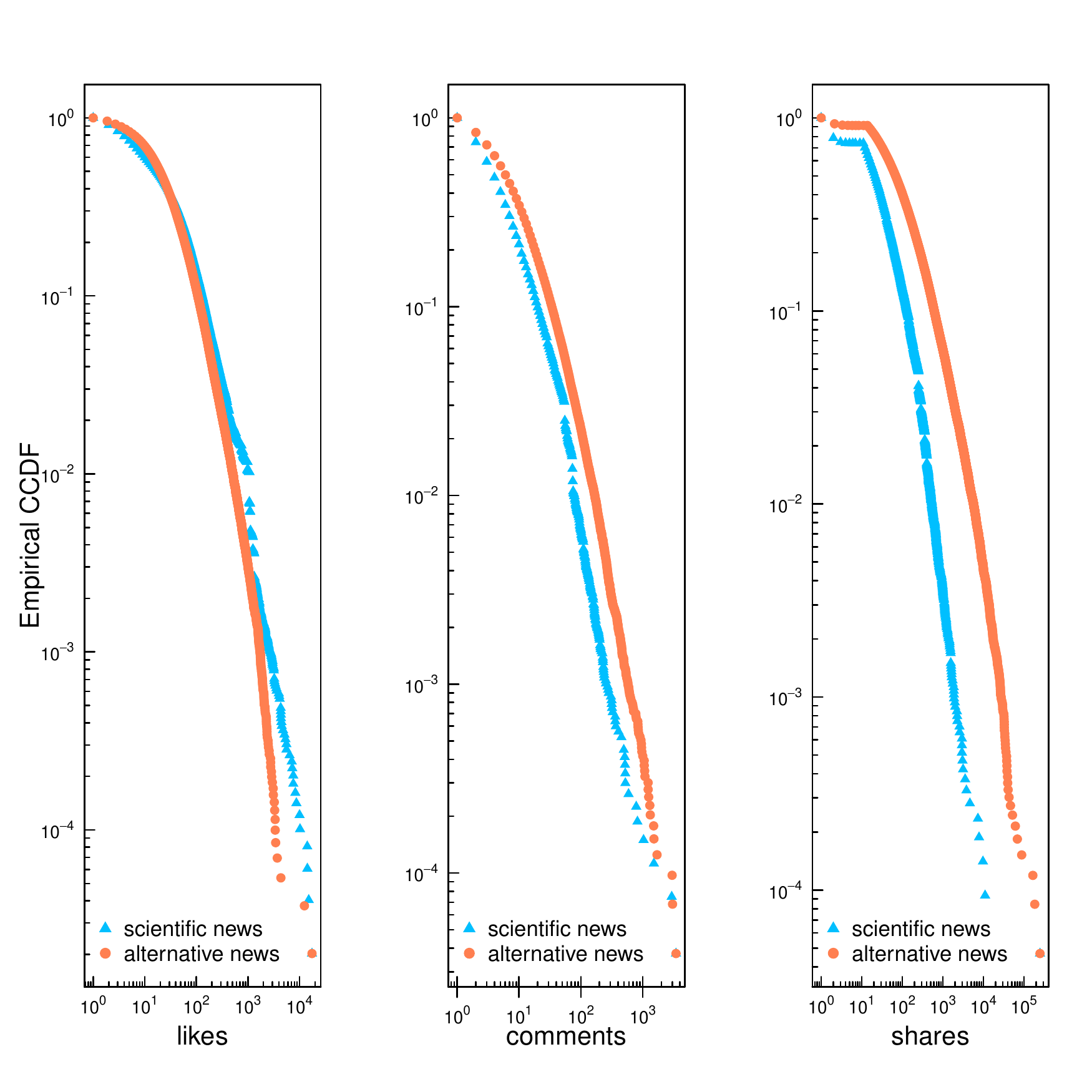}
\caption{\textbf{Users Activity}. Empirical complementary cumulative distribution function (CCDF) of users' activity (like, comment and share) for post grouped by page category.
The distributions are indicating similar consumption patterns for the various pages.}
\label{fig:activity_distribution}
\end{figure}

We start our analysis by looking at how Facebook users consume posts promoted by pages of alternative and mainstream scientific news.
In Figure \ref{fig:activity_distribution} we show the empirical complementary cumulative distribution function (CCDF) for likes (intended as positive feedbacks to the post), comments (a measure of the activity of online collective debates), and shares (intended as the attitude to share a given information) for all posts produced by the different categories of pages. 
Despite the very diverse nature of the information, posts are consumed in a similar way. 

\begin{figure}[H]
 \centering
       \includegraphics[width=.60\textwidth]{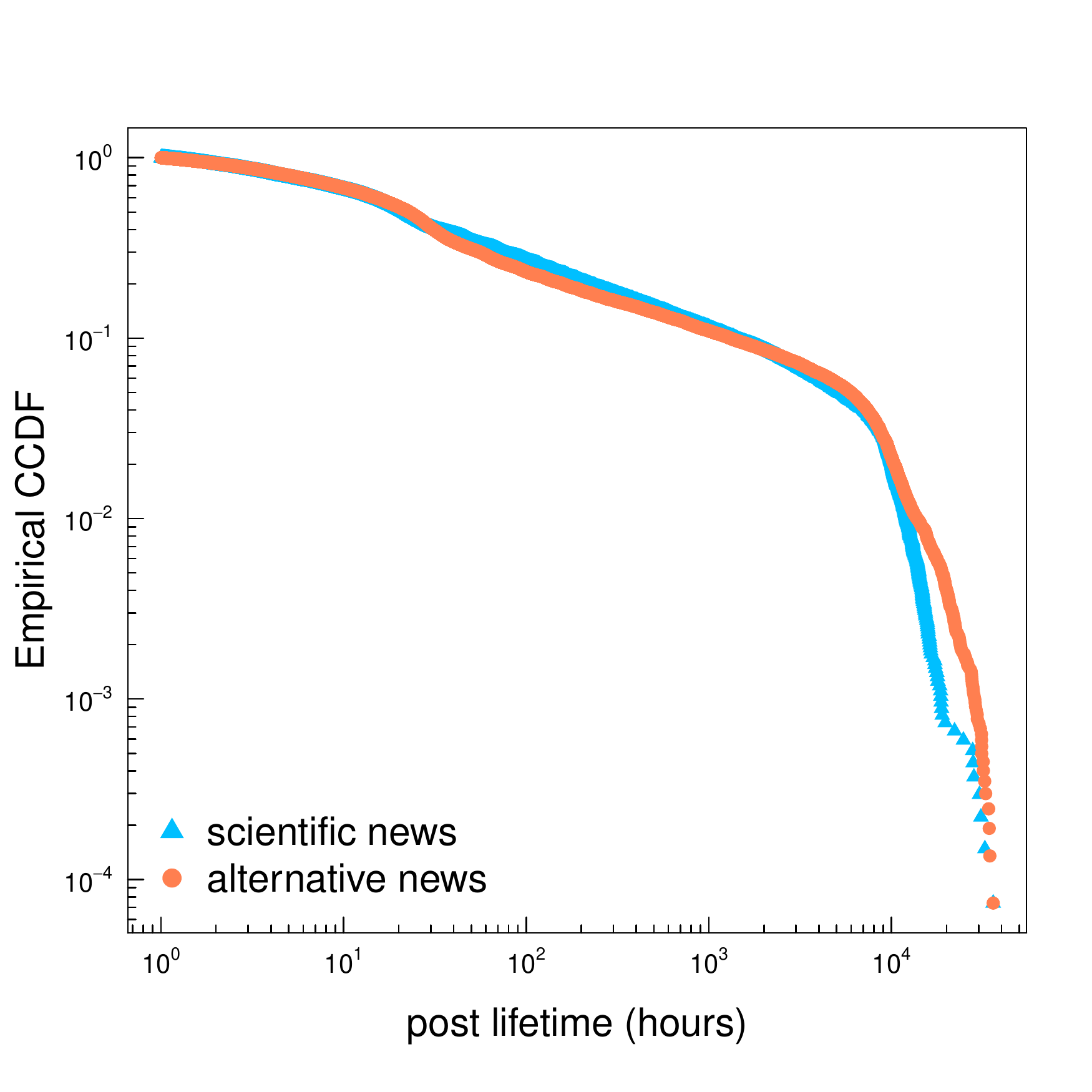}
\caption{\textbf{Post lifetime}. Empirical complementary cumulative distribution function (CCDF), grouped by page category, of the temporal distance between the first and last comment to each post. The life time of posts in both categories is similar.}
\label{fig:post_lifetime}
\end{figure}

A post sets the attention on a given topic and then a discussion may evolve in the form of comments to the post. 
Hence, to further investigate users consumption patterns, we zoom in at the level of comments.
In Figure \ref{fig:post_lifetime} we show CCDF of the posts lifetime -- i.e. the temporal distance between the first and the last comment for each post.
Such a measure is a good approximation of users attention to the information and in both case (scientific and alternative news) it has a comparable lifetime.

To account for the distinctive features of the communities' consumption patterns, we focus on the correlation of user's combination of interactions. 
In Table \ref{tab:data_corr} we show the Pearson correlation for user couple of actions on posts (likes, comments and shares). As an example, a high correlation coefficient for Comments/Shares indicates that posts more commented are likely to be shared and vice versa. 

\begin{center}
\begin{table}[!h]
\centering
\begin{tabular}{l|c|c|c|c}
\bf {  }  & \bf {Likes/Comments} & \bf {Likes/Shares} & \bf {Comments/Shares}  \\ \hline 
Science & $ 0.523 $ & $ 0.218 $ & $ 0.522 $ \\
Conspiracy & $ 0.639 $ & $\textbf{0.816} $ & $ 0.658 $ \\
\end{tabular}
  \caption{ \textbf{Users Actions.} Correlation (Pearson coefficient) between couple of actions to each post in scientific and alternative news. 
  Posts from alternative news pages are more likely to be liked and shared by users, indicating a major commitment in the diffusion.}
  \label{tab:data_corr}
\end{table}
\end{center}

Posts from alternative news have higher values than those on scientific news, and in particular they receive more likes and shares, indicating a strong attitude of users toward dissemination.
Such a result is consistent with \cite{Cinnirella_2013,Epstein96,Webster94} which state that conspiracists need for cognitive closure, i.e. they are more likely to interact with conspiracy based theories and have a lower trust in other information sources.

Qualitatively different information are consumed in the same way in terms of likes, comments and shares. Such a similarity of information consumption patterns is confirmed by the duration of online debates around a the various information. 
However, zooming in at the combination of actions, users of alternative news are more prone to share and like on a post. Such latter result indicates a higher level of commitment of consumers of conspiracy news. 

\subsection*{Detecting information-based communities}

The classification of pages in scientific and alternative (or conspiracy) related contents is grounded on their self-description and on the kind of diffused information. 
To validate such a partitioning we apply a network based approach aimed at measuring distinctive connectivity patterns of these information-based communities -- i.e. users consuming information belonging to similar narratives.

We consider a bipartite network having as nodes users and affiliation the Facebook pages. 
A comment to a given information posted by a page determines a link between a user and a page.

More formally, a bipartite graph is a triple $\mathcal{G}=(A,B,E)$ where $A=\left\{ a_{i}\,|\,i=1\dots n_{A}\right\} $ and $B=\left\{ b_{j}\,|\,j=1\dots n_{B}\right\} $ are two disjoint sets of vertices, and $E\subseteq A\times B$ is the set of edges -- i.e. edges exist only between vertices of the two different sets $A$ and $B$. The bipartite graph $\mathcal{G}$ is described by the matrix $M$
defined as

\[
M_{ij}=\left\{ \begin{array}{cc}
1 & if\, an\, edge\, exists\, between\, a_{i}\, and\, b_{j}\\
0 & otherwise
\end{array}\right.
\]

For our analysis we use the co-occurrence matrices $C^{A}=MM^{T}$
and $C^{B}=M^{T}M$ that count, respectively, the number of common neighbors between two vertices of $A$ or $B$. 
$C^{A}$ is the weighted adjacency matrix of the co-occurrence graph $\mathcal{C}^{A}$ with vertices on $A$. Each non-zero element of $C^{A}$ corresponds to an edge among vertices $a_{i}$ and $a_{j}$ with weight $P_{ij}^{A}$.

\begin{figure}[H]
 \centering
       \includegraphics[width=.95\textwidth]{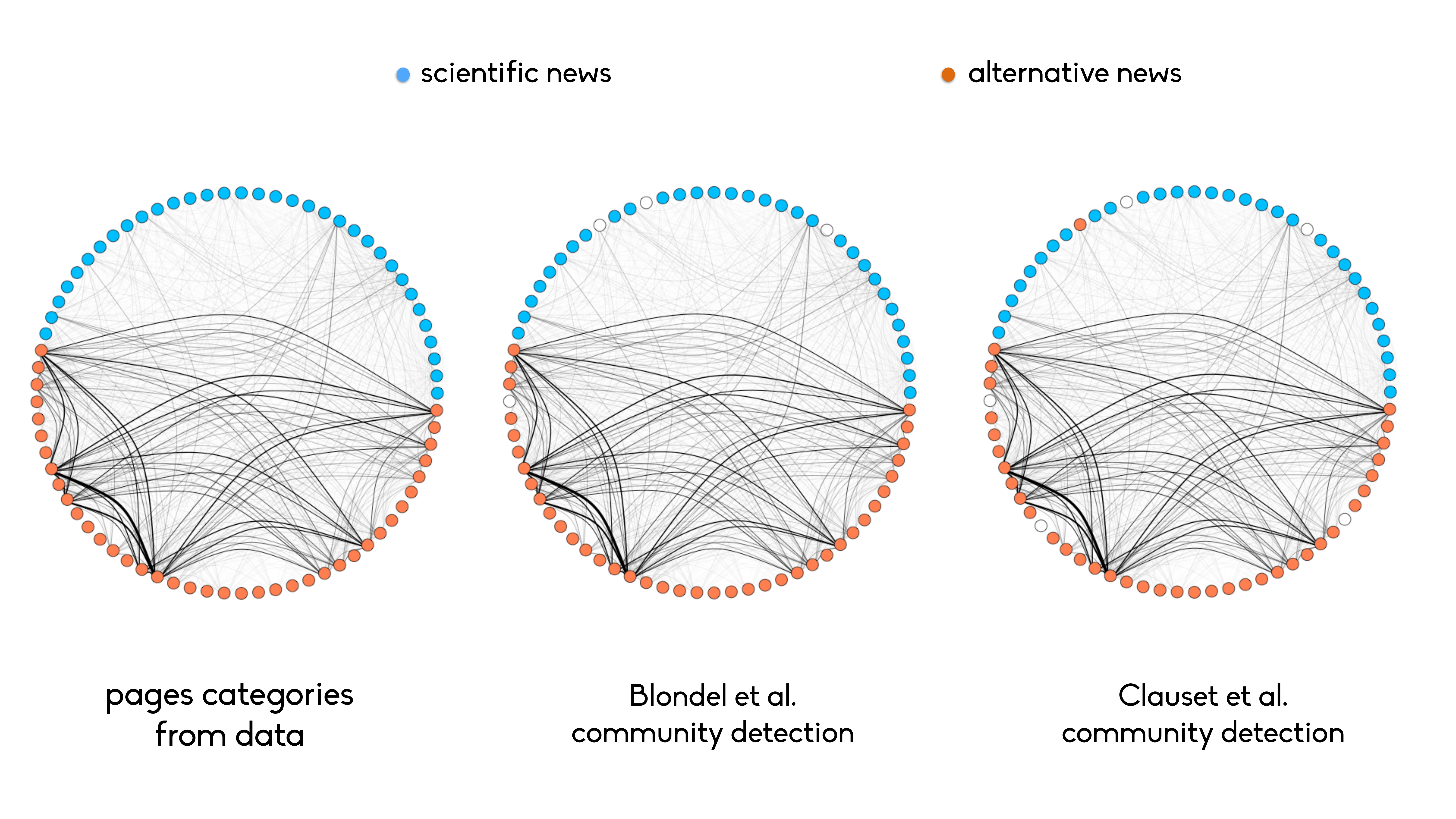}
\caption{\textbf{Page Network.}  
The membership of 73 pages as a) identified by means of their self-description, b) by applying the multi-level modularity optimization algorithm, and c) by looking at the maximum modularity score. Community detection algorithms based on modularity are good discriminants for the community partitioning.}
\label{fig:communities_detection}
\end{figure}

To test the community partitioning we use two well known community detection algorithms based on modularity \cite{Blondel2008,Clauset2004}. 
The former algorithm is based on multi-level modularity optimization. 
Initially, each vertex is assigned to a community on its own. In every step, vertices are re-assigned to communities in a local, greedy way. 
Nodes are moved to the community in which they achieve the highest modularity. 
Differently, the latter algorithm looks for the maximum modularity score by considering all possible community structures in the network.  We apply both algorithms to the bipartite projection on pages. In Figure \ref{fig:communities_detection} we show the membership of pages as a) identified by means of their self-description, b) by applying the multi-level modularity optimization algorithm and c) by looking at the maximum modularity score.
The detected membership identifies nodes of the categories with a good approximation.

The connectivity patterns, in particular the modularity, are good indicators for the community structure. Since we are considering users' comments on the pages' posts, this aspect is pointing out a more dense interaction among users of the same community on the conspiracy side.

\subsection*{Polarized users and their interaction patterns }

In this section we want to understand if it is possible to identify interaction patterns of communities  with respect to the information both in favor and against their system of beliefs. 
Hence, we zoom in at the level of communities by labeling users through a simple thresholding algorithm accounting for the percentage of likes on one or the other category. 
The choice of the \emph{like} as a discriminant is grounded on the fact that generally such an action stands for a positive feedback to a post \cite{Viswanath:2009}. 
Therefore we will consider user to be polarized in a community when the number of his/her likes with respect to his/her total like activity on one category -- scientific or conspiracy news -- is higher than 95\%.

More formally, the labeling algorithm can be described as thresholding strategy on the total number of users likes. Considering the total number of likes of a user $L_u$ on both posts $P$ in categories $S$ and $C$. Let  $l_s$ and $l_c$ define the number of likes of a user $u$ on $P_s$ or $P_c$, respectively denoting posts from scientific and conspiracy pages.  Then, we will have the total like activity of users on one category expressed as $\frac{l_s}{L_u}$. Fixing a threshold $\theta$ we can discriminate users with enough activity on one category. More precisely, the condition for a user to be labeled as a polarized user in one category can be described as $\frac{l_s}{L_u}$ $\vee$ $\frac{l_c}{L_u}$ $> \theta$.

For our analysis we fixed $\theta$ to 0.95, hence we will consider a user to be polarized on one category if he has at least the 95\% of likes there. 
Our thresholding algorithm applied to our Facebook data is able to identify 255,225 polarized users of scientific pages (76,79\% of the users that interacted on scientific pages in terms of liking) and 790,899 conspiracy polarized users (91,53\% of the users that interacted on conspiracy pages in terms of liking).

Now, if we look at commenting activity of polarized users inside and outside their community, we notice that users polarized on conspiracy pages tend to interact especially in their community both in terms of comments (99,08\%) and likes.
Conversely, users polarized in science seem to act slightly more outside their community (90,29\%). The results are shown in Table \ref{tab:polarized_activity}.

\begin{center}
\begin{table}[H]
\centering
\begin{tabular}{l|c|c|c|c|c}
\bf {  }  & \bf {Users} & \bf {(\%) Users} & \bf {Comments on} & \bf {Comments on the} & \bf {Comments on} \\
\bf {  }  & \bf {classified} & \bf {classified} & \bf {their category} & \bf {opposite category} & \bf {both categories}\\ 
\hline 
Scientific News & $ 255,225 $ & $ 76,79 $ & $ 126,454 $ & $13,603$ & $ 140,057 $\\
Alternative News & $ 790,899 $ & $ 91,53$ & $ 642,229 $ & $5,954$& $ 648,183 $\\
\end{tabular}
  \caption{\textbf{Activity of polarized users.} Number of classified users for each category and their commenting activity on the category in which they are classified and on the opposite category. Users polarized on conspiracy pages tend to interact especially in their community both in terms of comments and likes. Users polarized in science are more active elsewhere.}
  \label{tab:polarized_activity}
\end{table}
\end{center}

\begin{figure}[H]
 \centering
       \includegraphics[width=.65\textwidth]{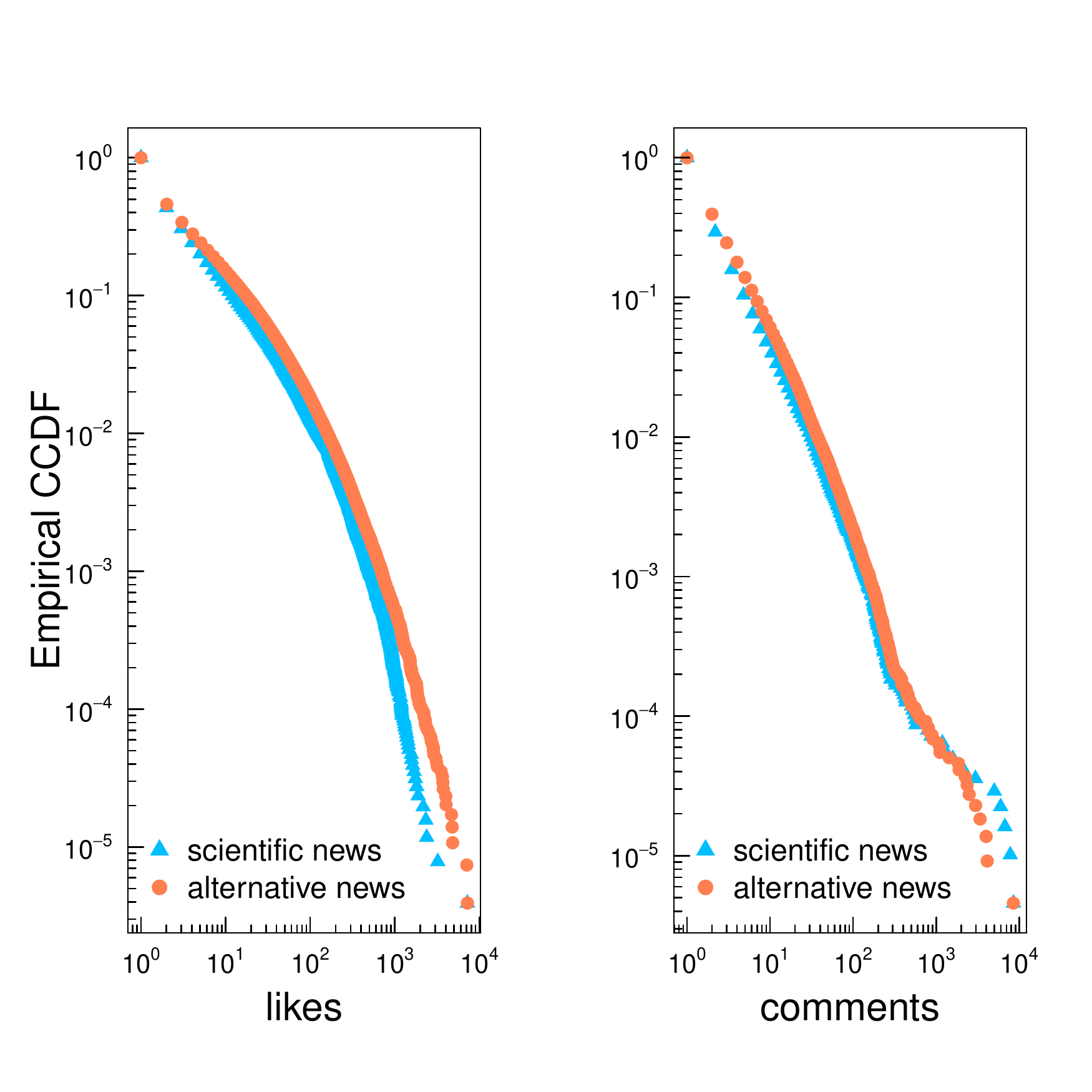} 

\caption{\textbf{Consumption patterns of polarized users.} Empirical complementary cumulative distribution function (CCDF) for likes and comments of polarized users.}
\label{fig:activity_polarized_distribution}
\end{figure}

Figure \ref{fig:activity_polarized_distribution} shows the CCDF for likes and comments of polarized users. 
Despite the very profound different nature of the information, the consumption patterns are nearly the same both in terms of likes and comments, indicating that different communities based on very different information behave in a similar way.

Summarizing, from these results we observe that, as expected, alternative and scientific news aggregate individuals in well separated communities. 
Furthermore, polarized users of alternative and scientific news consume information in a similar way.

\begin{figure}[H]
 \centering
       \includegraphics[width=.75\textwidth]{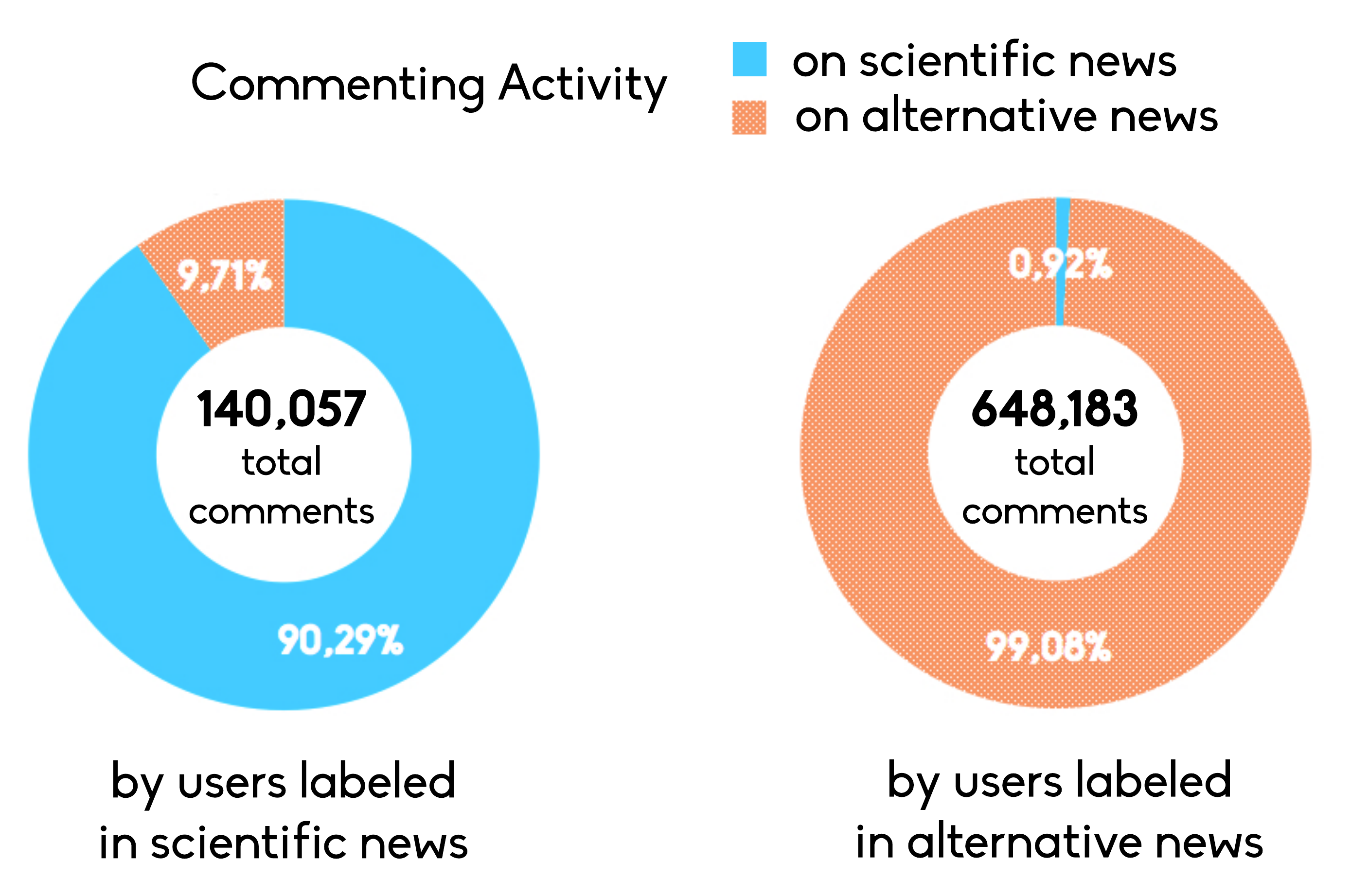}
\caption{\textbf{Activity between communities}.  
Posts on which at least a member of each the two communities has commented. The number of posts is 7751 (1991 from scientific news and 5760 from alternative news). 
Here we show the commenting activity in terms of polarized users on the two categories.}
\label{fig:act_between_communities}
\end{figure}

As a further investigation, we select the set of posts on which at least a member of each the two communities has commented.
We find polarized users of communities debating on 7751 posts (1991 from scientific news and 5760 from alternative news). 
We look at the activity of 95\%-threshold polarized users on the opposite topic in terms of comments. As shown in Figure \ref{fig:act_between_communities}, polarized users of scientific news made 13,603 comments on post published by alternative news (9.71\% of their total commenting activity), whereas polarized users of alternative news commented on scientific posts only 5,954 times (0.92\% of their total commenting activity, i.e. roughly ten times less than polarized users of scientific news). 

Usual followers of alternative news are more prone to interact inside their community. In turn, followers of scientific news are more likely to comment with the aim to inhibit the diffusion of \textit{unsubstantiated} claims coming from conspiracy pages.

\subsection*{Response to false information}

In \cite{Mocanu2014} we noticed that people who engage with debates on conspiracy news posts are much more likely to engage in false news posted by trolls. 
Conspiracy theories seem to come about by a process in which ordinary satirical commentary or obviously false content somehow jump the credulity barrier, mainly because of the unsubstantiated nature of conspiracy related information.
Therefore, we want to check how polarized users of opposite narrative interact with posts that are deliberately false.

We collected a set of troll posts -- i.e. paradoxical imitations of alternative information sources. 
These posts are clearly unsubstantiated claims, like the undisclosed news that infinite energy has been finally discovered, or that a new lamp made of actinides (e.g. plutonium and uranium) might solve problems of energy gathering with less impact on the environment, or that the chemical analysis revealed that chem-trails contains sildenafil citratum (the active ingredient of Viagra\textsuperscript{TM}).

\begin{figure}[H]
 \centering
       \includegraphics[width=.75\textwidth]{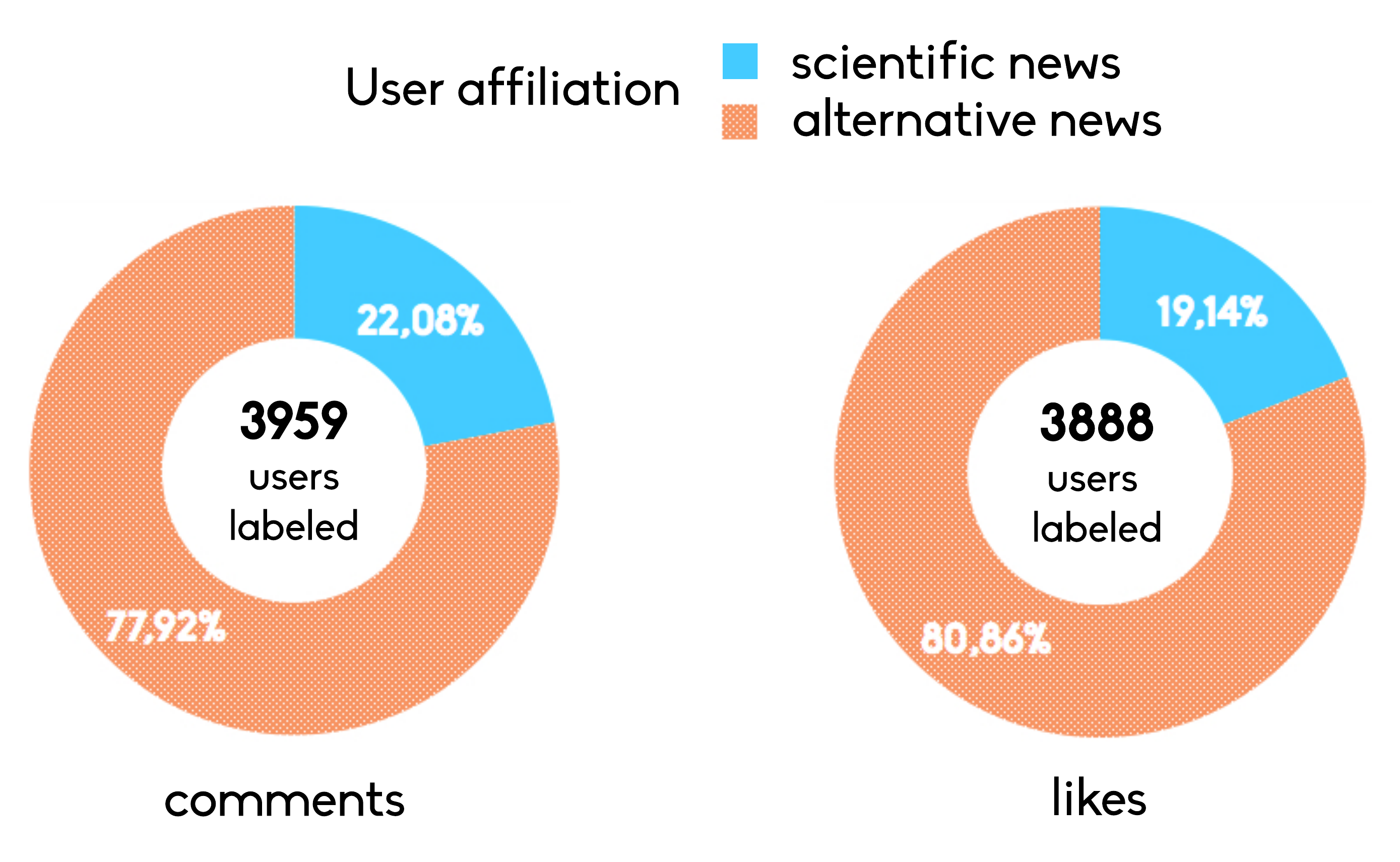}
\caption{\textbf{Polarized users on false information.}  
Percentage of comments and likes on intentional false memes posted by a satirical page from polarized users of the two categories.}
\label{fig:activity_on_troll}
\end{figure}

Figure \ref{fig:activity_on_troll} shows how polarized users of both categories interact with troll posts in terms of comments and likes. 
On this external sample, coherently with our previous results, we find that polarized users of both categories retain their interaction patterns. 
Polarized users of conspiracy pages are more active in liking and commenting.

\section*{Conclusions}
Recently in \cite{Mocanu2014} has been shown that unsubstantiated claims reverberate for a timespan comparable to the one of more verified information and that usual consumers of conspiracy theories are more prone to interact with them. 
Conspiracy theories, in particular, find on the internet a natural medium for their diffusion and, not rarely, trigger collective counter-conspirational actions \cite{Atran2012,Lewandowsky2013}.  
Narratives grounded on conspiracy theories tend to reduce the complexity of reality and are able to contain the uncertainty they generate \cite{byford2011conspiracy,finerumor,hogg2011extremism}.

Along this path, in this work we studied how users interact with information related to different (opposite) narratives on Facebook.
Through a thresholding algorithm we label polarized users on the two categories of pages identifying well shaped communities. 
In particular, we measure commenting activity of polarized users on the opposite category, finding that polarized users of conspiracy news are more focused on posts of their community and their interaction behavior is more oriented to the diffusion of news.
On the other hand, polarized users of scientific news are less committed in the diffusion and more prone to comment on conspiracy pages. 
A possible explanation for such behavior is that the former want to diffuse what is neglected by main stream thinking, whereas the latter aims at inhibiting the diffusion of conspiracy news and proliferation of narratives based on unsubstantiated claims.  
Finally, we test how polarized users of both categories responded to the inoculation of 4709 false claims produced by a parodistic page, finding polarized users of conspiracy pages to be the most active.

We have shown that communities tend to organize according to shared narratives. Our quantitative observations open interesting questions on the role of social reinforcement in shaping communities attitude and resistance to external beliefs. 
However, according to the preference with the kind of information sources, we are able to recognize users more prone to trigger viral attention on a given topic and, in turn, such a result open to the possibility of designing ad-hoc communication strategy to contrast and immunize unsubstantiated narratives.
These results are coherent with the findings of\cite{Cinnirella_2013,Epstein96,Webster94} indicating the existence of a relationship between beliefs in conspiracy theories and the need for cognitive closure. Those who use a more heuristic approach when evaluating evidences to form their opinions are more likely to end up with an account more consistent with their existing system of beliefs. However, anti-conspiracy theorists may not only reject evidence that points toward a conspiracy theory account, but also spend cognitive resources for seeking out evidences to debunk conspiracy theories even when these are satirical imitation of false claims.

\section*{Acknowledgments}

Funding for this work was provided by EU FET project MULTIPLEX nr. 317532 and SIMPOL nr. 610704. The funders had no role in study design, data collection and analysis, decision to publish, or preparation of the manuscript.
Special thanks go to Delia Mocanu, "Protesi di Protesi di Complotto", "Che vuol dire reale", "La menzogna diventa verita e passa alla storia", "Simply Humans", "Semplicemente me", Salvatore Previti, Brian Keegan, Dino Ballerini, Elio Gabalo, Sandro Forgione and "The rooster on the trash" for their precious suggestions and discussions.

\bibliography{trolling}

\end{document}